 \documentclass[
  pre,
  aps,
  a4paper,
  english,
  reprint,
  twocolumn,
  showpacs,
  superscriptaddress
 ]{revtex4-1}

\usepackage{graphicx}
\usepackage{color}
\usepackage{bm}
\usepackage{amsfonts}
\usepackage{amsmath,amssymb}
\usepackage{hyperref}
\usepackage{soul}
\usepackage[mathscr]{euscript}
\usepackage{physics}
\usepackage{dcolumn}
\usepackage{subfigure}
\usepackage{adjustbox}
\newtheorem{theorem}{Theorem}

\begin{document}

\title{A Mathematical Framework for Misinformation Propagation in Complex Networks: Topology-Dependent Distortion and Control}

\author{Saikat Sur}
\email{saikats@imsc.res.in}
\affiliation{Optics \& Quantum Information Group, The Institute of Mathematical Sciences,
HBNI, CIT Campus, Taramani, Chennai 600113, India}
\affiliation{Department of Chemical and Biological Physics \& AMOS,
Weizmann Institute of Science, Rehovot 7610001, Israel}

\author{Rohitashwa Chattopadhyay}
\email{rohitashwa.chattopadhyay@tifr.res.in}
\affiliation{Department of Theoretical Physics, Tata Institute of Fundamental Research,
Homi Bhabha Road, Mumbai 400005, India}

\author{Jens Christian Claussen}
\email{j.c.claussen@bham.ac.uk}
\affiliation{School of Computer Science, University of Birmingham, Edgbaston, B15 2TT, United Kingdom}

\author{Archan Mukhopadhyay$^{\dagger}$}
\email{archanmukhopadhyay.pi.ns@msruas.ac.in}
\affiliation{Department of Physics, M. S. Ramaiah University of Applied Sciences, Bengaluru 560058, India}

\thanks{$^{\dagger}$Corresponding author.}

\date{\today}
\begin{abstract}
Misinformation is pervasive in natural, biological, social, and engineered
systems, yet its quantitative characterization remains challenging due to
context-dependent errors and the heterogeneous structure of real-world
interaction networks. We develop a general mathematical framework for
quantifying information distortion in distributed systems by modeling how
local transmission errors accumulate along network geodesics and reshape
each agent's perceived global state. Through a drift--fluctuation
decomposition of pathwise binomial noise, we derive closed-form expressions
for node-level perception distributions and show that directional bias
induces only a uniform shift in the mean, preserving the fluctuation
structure. This establishes a previously unreported shift-invariance
principle governing error propagation in networks. Applying the framework
to canonical graph ensembles, we uncover strong topological signatures of
misinformation: Erd\H{o}s--R\'enyi random graphs exhibit a double-peaked
distortion profile driven by connectivity transitions and geodesic-length
fluctuations, scale-free networks suppress misinformation through
hub-mediated integration, and optimally rewired small-world networks achieve
comparable suppression by balancing clustering with short paths. A direct
comparison across regular lattices, Erd\H{o}s--R\'enyi random graphs,
Watts--Strogatz small-world networks, and Barab\'asi--Albert scale-free
networks reveals a connectivity-dependent crossover. In the extremely sparse
regime, scale-free and Erd\H{o}s--R\'enyi networks behave similarly. At
intermediate sparsity, Watts--Strogatz small-world networks exhibit the
lowest misinformation. In contrast, Barab\'asi--Albert scale-free networks
maintain low misinformation in sparse and dense regimes, while regular
lattices produce the highest distortion across connectivities. We
additionally show how sparsity constraints, structural organization, and
connection costs delineate regimes of minimal misinformation. Overall, our
results provide an analytically tractable foundation for understanding and
controlling information reliability in complex networked systems.
\end{abstract}
\maketitle

\section{Introduction}

A coordinated response in a multi-agent system relies on each agent having 
an accurate estimate of the global state. In real settings, however, such 
accuracy is rarely achieved. Physical, biological, social, and 
environmental sources of noise introduce small but unavoidable errors, and 
understanding how these errors propagate through an interaction network is 
essential, especially because collective performance can be extremely 
sensitive to local inaccuracies. In this work, the term \emph{misinformation} 
refers precisely to the deviation between the true global state and the 
state perceived at each agent, arising from imperfect propagation across 
the network rather than from Shannon-theoretic information measures. {\color{black} Such 
errors may originate unintentionally from stochastic processes or 
intentionally from adversarial influences, yet their consequence on 
collective reliability is similar as they distort how individuals in the system perceive 
others.}

{\color{black} Quantification and dynamics of misinformation have been studied across journalism, sociology, neuroscience, 
collective biological systems, distributed computing, and networked cyber-physical 
systems,} but a unified mathematical understanding remains limited. Prior 
studies have offered domain-specific or simulation-driven insights, such as 
qualitative analyses of error propagation based on connectivity measures 
\cite{Krol_Kukla_2012}, error accumulation in deep neural architectures 
\cite{Li2017}, and preferred interaction topologies in sociotechnical and 
biological settings \cite{klemm2005}. {\color{black} Efforts to restrict misinformation on 
online platforms \cite{Ghoshal2025} and to characterize the spread of true 
versus false information in complex networks \cite{NIAN2021} provide 
valuable perspectives,  but do not yield a general analytical  framework for 
quantifying misinformation in distributed systems.} \textcolor{black}{Complex networks display emergent macroscopic behavior and are vulnerable to localized failures and  perturbations that can propagate through the interconnected structure, leading to dysfunction. Recent theoretical and computational frameworks characterize these effects in terms of robustness and resilience, thus providing tools to analyze, predict, and mitigate network breakdown~\cite{ji_pr_2023,artime_nrp_2024}.}

Information flow in any multi-agent system is fundamentally shaped by  
 connection geometry of the network. Signals often traverse multiple 
intermediaries, meaning that the state perceived by an agent is a non-local 
function of the network topology \cite{newman_2010,barabasi2016}. 
Consequently, errors accumulate along geodesic paths, as seen in 
network-driven propagation phenomena \cite{pastor2001,brockmann2013}. The 
efficiency of information transfer therefore depends critically on the 
distribution of path lengths and on the small-world properties of the 
network \cite{watts_nature_1998}. When information is represented as a 
real-valued field, biased errors introduce additional randomness, implying 
that a complete description must incorporate both intrinsic fluctuations and 
topological constraints.

{\color{black} In this article, we develop a general mathematical framework for quantifying
misinformation by modeling the state propagation as a path-accumulated
error process. Our framework is built on the assumption that each agent has
only local access to information---that is, an agent can perceive the global state of the network
solely through the states of other nodes as they arrive via error-prone paths.} A central analytic result is a drift--fluctuation 
decomposition showing that the bias parameter influences perception 
purely through a uniform translation of the mean, while the fluctuation 
structure remains invariant. This shift-invariance property, previously 
unidentified in network error-propagation models, greatly simplifies the 
resulting misinformation measure and enables closed-form expressions for the 
Gaussian and delta-function limits.

Applying the framework to canonical network ensembles reveals strong and 
previously unreported structural effects. Most strikingly, Erd\H{o}s--R\'enyi 
networks exhibit a non-monotonic misinformation profile with two 
well-separated local maxima, arising from the interplay between the 
emergence of the giant component, the distribution of geodesic distances, 
and the formation of short cycles. To our knowledge, this double-peaked 
behavior has not been documented in prior studies of information distortion 
on networks. In contrast, scale-free networks suppress misinformation 
through hub-mediated global coordination, whereas optimally rewired 
small-world networks suppress it by balancing strong clustering with short 
characteristic path lengths, consistent with rewiring effects reported in 
\cite{klemn2002}. {\color{black}These distinct but complementary mechanisms may offer 
insights into the reason behind the widespread existence of hub-dominated and small-world architectures  
in natural and engineered information-processing systems.}

{\color{black} Overall, our work introduces an  analytically tractable framework 
for measuring misinformation and identifying network topologies that 
optimize collective reliability. This framework integrates several parameters, viz., error accumulation via geodesics, network 
geometry, and collective perception; thereby elucidates how structures shape information fidelity in complex systems, with implications 
for biological networks, distributed sensing, communication systems, and 
socio-technical platforms.}

\section{Noisy Information Flow in Biological Networks}



{\color{black} Biological networks provide natural examples in which information flow is 
inherently noisy. This noise might often have important functional consequences.} Chemotactic signaling in 
\emph{E.\ coli} fluctuates due to stochastic ligand--receptor binding and 
downstream processing \cite{berg2004coli,tu2008modeling}; synaptic 
transmission is probabilistic and exhibits variability in spike timing 
\cite{faisal2008noise,dayan2001theoretical}; gene regulatory networks show 
intrinsic noise that influences cell-fate decisions 
\cite{elowitz_nature_2000,elowitz2002stochastic,raj2008nature}; and immune 
and protein--interaction networks display variability and crosstalk that 
reshape collective responses \cite{paul2012fundamental,barabasi2004network}. 
{\color{black}These systems highlight that misinformation generated by noise is 
unavoidable, and amplification, buffering, or redirection of such errors is governed by the structure of the network.}

{\color{black} Neuronal networks are another example of path-induced errors. Neurons communicate via 
action potentials that propagate across synapses, with spike timing and 
firing rates encoding sensory and cognitive information. Synaptic 
transmission, however, is inherently noisy as spikes may fail, arrive with 
temporal jitter, or be distorted by background activities. Such variability might prohibit coding fidelity and lead to 
pathological synchrony such as epileptic seizures~\cite{faisal2008noise,
dayan2001theoretical,destexhe2003neuronal}.}

Gene regulatory networks also illustrate how misinformation can arise from 
intrinsic stochasticity. Transcription factors regulate gene expression 
through activation and repression, but low copy numbers of mRNA and 
proteins introduce significant fluctuations. This noise can misguide 
cell-fate decisions or perturb developmental pathways~\cite{elowitz2002stochastic,
raj2008nature,alon2007introduction}. Network motifs such as feed-forward 
loops and negative-feedback circuits are thought to buffer against such 
stochasticity.

The immune system offers a further example. Receptor signaling, cytokine 
communication, and transcriptional responses form a distributed network 
that are used to discriminate self from non-self. Noisy or ambiguous signals 
may be amplified inappropriately, generating autoimmunity, while legitimate 
pathogen cues may fail to elicit a sufficient response~\cite{paul2012fundamental,
perelson1997immune,medzhitov1997,hoffmann2006immune}.

Finally, protein--protein interaction (PPI) networks transmit signals via 
cascades of binding interactions. Non-specific binding, competitive 
interactions, and pathway crosstalk can misroute signals. The scale-free 
organization of many PPI networks provides resilience against random 
failures but also creates vulnerabilities due to reliance on highly 
connected hubs~\cite{barabasi2004network,han2004evidence,jeong_nature_2001}.

{\color{black}To sum up, the above examples demonstrate that misinformation due to noise is 
not merely a theoretical construct but a challenge that pervades across biology. 
They highlight how network topology shapes the reliability of distributed 
information processing. This reinforce the necessity for a general 
framework that captures how structures amplify or 
suppress errors across complex systems, especially across complex networks \cite{CLAUSSEN2007,KIM2008}}.

\section{Model}
\label{sec:2}
Let us consider a connected network $\mathcal{G}$ consisting of $n$ nodes (or agents) connected through $m$ links. The information stored at each node is represented as a  multi-dimensional real vector, describing the state of that node. Our goal is to formulate a field over the network---namely, the collection of these  multi-dimensional states as a function of node indices. For expository clarity, we restrict attention to a one-dimensional
scalar field.  A higher-dimensional vector field can always be
decomposed into a collection of independent orthogonal scalar
components, each of which can be treated separately within the
same framework. Thus, this simplification does not reduce the scientific generality of the model---it merely allows for a cleaner presentation.

We denote the state of node $i$ by $X_i$. The ensemble of node states $\{X_i\}_{i=1}^n$ is described by a random variable $\mathbf{X}$, which takes real values $x$ distributed according to a probability density function (PDF) $P(x)$. To make the model analytically tractable, we assume that the state of a node can influence or propagate to another node exclusively through the shortest path connecting them. This assumption effectively constrains information flow along the shortest paths (geodesic distances) on $\mathcal{G}$, providing a well-defined structure for modeling dynamical or diffusion-like processes on the network.

\begin{figure*}[t]
    \centering
    \includegraphics[width=0.85\linewidth]{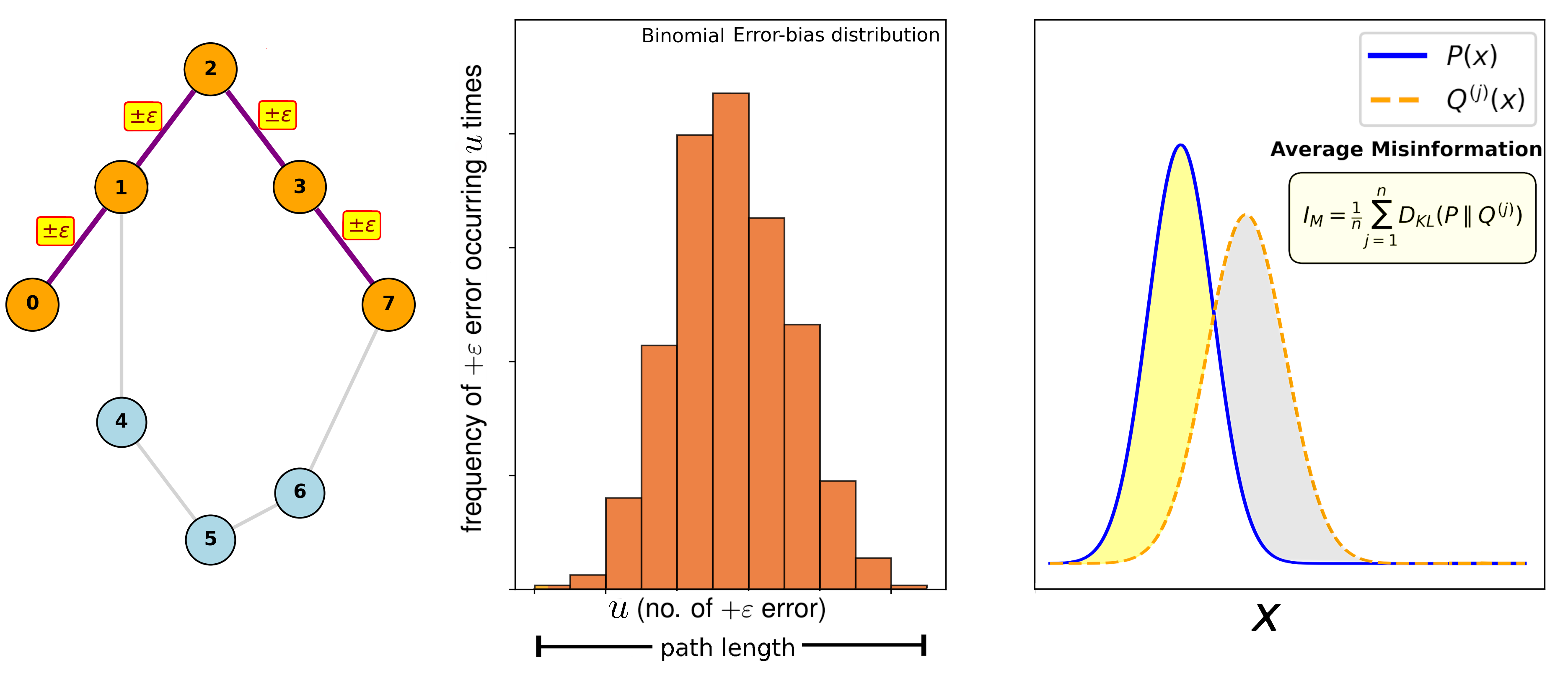}
    \caption{
    \textbf{Schematic of the misinformation model.} \textcolor{black}{\textit{Left:} Illustration of error accumulation between a pair of agents in a network, where each interaction along the shortest path contributes a random error of size $\pm \varepsilon$. \textit{Middle:} The binomial distribution of the accumulated errors. \textit{Right:} Comparison between the true distribution $P(x)$ and the perceived distributions $Q^{(j)}(x)$, with the average misinformation quantified by the node-averaged KL divergence.}
    }
    \label{fig:misinfo_model}
\end{figure*}

{\color{black}
Let us denote by $X^{(j \leftarrow i)}$ the information that the $j$th node perceives about the $i$th node, and by $d_{ij}$ the shortest path connecting the pair $(i,j)$. To model the bias affecting the error---whether positive or negative---we draw from a binomial distribution, where the number of trials equals the shortest path length between the nodes. Thus, the error introduced along the path consists of $d_{ij}$ repeated independent Bernoulli trials. In each trial, the probability of a positive error $(+\varepsilon)$ is $r$, and the probability of a negative error $(-\varepsilon)$ is $(1-r)$. We normalize the magnitude of the error per unit path by the largest possible geodesic distance between any two nodes, which is $(n-1)$. Accordingly, we set the magnitude of the error per unit path to be
\begin{equation}
|\varepsilon| = \frac{1}{n-1}.
\end{equation}
Since $d_{ij} \le (n-1)$ for all $(i,j)$, this normalization ensures that the maximum cumulative error is bounded by
\begin{equation}
\bigl|X^{(j \leftarrow i)} - X_i \bigr| \le 1.
\end{equation}
Let $u$ denote the number of times a positive error occurs along the $d_{ij}$ trials; then $(d_{ij}-u)$ denotes the number of negative errors. The perceived value of $X_i$ at node $j$ is therefore
\begin{eqnarray}
X^{(j \leftarrow i)}
&=& X_i + \frac{1}{n-1}\left[u - (d_{ij} - u) \right]
\nonumber \\
&=& X_i + \frac{1}{n-1}\left( 2u - d_{ij} \right).
\label{eq:perceived}
\end{eqnarray}
In our protocol, as we traverse each unit path (i.e., link), we select the error sign from a binomial distribution. The probability that exactly $u$ positive errors occur in $d_{ij}$ independent trials is
\begin{equation}
S(d_{ij}, r, u)
= {d_{ij} \choose u} r^u (1-r)^{d_{ij}-u},
\label{eq:binomial}
\end{equation}
where $u \in [0,d_{ij}]$. The maximum deviation in the perceived state $X^{(j \leftarrow i)}$ occurs at the extreme cases $u=0$ or $u=d_{ij}$, in which
\begin{equation}
\bigl|X^{(j \leftarrow i)} - X_i \bigr| = \frac{d_{ij}}{n-1}.
\end{equation}
On the other hand, the error is completely nullified when $u=d_{ij}/2$, provided $d_{ij}$ is even. {{\color{black}(please refer to the left and the middle panel
of Fig.~\ref{fig:misinfo_model} for a schematic representation of our model)}}.
}
\subsection{Quantifying Misinformation}
In reality, the global state of the network is encoded in the random variable $\mathbf{X}$. However, because errors are introduced during propagation, this global information is perceived at the $j$th node as a different random variable, denoted $\mathbf{X}_j$. Consequently, the original PDF\ $P(x)$ is viewed at node $j$ as a modified distribution $Q^{(j)}(x)$. The misinformation encountered at node $j$ can therefore be quantified by measuring the relative distance between the true distribution $P(x)$ and the perceived distribution $Q^{(j)}(x)$. A natural and widely used metric for this purpose is the relative entropy, or Kullback--Leibler (KL) divergence~\cite{desurvire_2009, cover_2006}. The KL divergence between $P(x)$ (true) and $Q^{(j)}(x)$ (perceived) is defined as
\begin{eqnarray}
D_{KL}\!\left(P \,\Vert\, Q^{(j)}\right)
&=& \sum_{x} P(x)\, \ln \frac{P(x)}{Q^{(j)}(x)}
\nonumber \\
&=& H\!\left(P, Q^{(j)}\right) - H\!\left(P\right),
\label{eq:D}
\end{eqnarray}
where \(H(P)\) is the Shannon entropy of $P(x)$ and \(H(P, Q^{(j)})\) is the 
cross-entropy that measures the coding cost of $P(x)$ under $Q^{(j)}(x)$~\cite{cover_2006,mackay_2003}. {\color{black}(please refer to
the right panel of Fig.~\ref{fig:misinfo_model})}.

To characterize the overall distortion experienced by the system, we compute the average misinformation across all nodes. The misinformation per node is defined as
\begin{eqnarray}
I_M
&=& \frac{1}{n}\sum_{j=1}^{n} D_{KL}(P \Vert Q^{(j)}) \nonumber \\
&=& -\frac{1}{n} \sum_{j=1}^{n} \sum_{x} P(x) \ln Q^{(j)}(x) 
    - H(P).
\label{eq:I_M}
\end{eqnarray}

This quantity depends on both the topology of the network $\mathcal{G}$ and the bias parameter $r$ in the binomial error model. From Eq.~\ref{eq:D}, it is evident that misinformation reaches its maximum when the sum of cross-entropies across all nodes is maximized.


 {\color{black} The perceived distributions $Q^{(j)}(x)$ deviate from the true distribution $P(x)$ due to the accumulation of errors. In the limit of large networks ($n \rightarrow \infty$), the distributions become continuous and the summations are replaced by integrals


\begin{eqnarray}
I_M
&=&  -\frac{1}{n}\sum_{j=1}^{n} \int dx\,  P(x)\,\ln\!\left(1+\frac{Q^{(j)}(x) - P(x)}{P(x)}\right).
\label{eq:continuous_I}
\end{eqnarray}


In this article, however, we ultimately focus on cases where the true state distribution \(P(x)\) is sharply localized and approaches a Dirac delta distribution. This limit corresponds to systems in which the global state is precisely defined and all misinformation arises strictly from propagation errors rather than intrinsic variability. Before analyzing this limit, we illustrate Eq.~\ref{eq:continuous_I} for a smooth distribution. {\color{black}For this purpose, we consider a Gaussian 
\begin{equation}
P(x) =\frac{1}{\sqrt{2\pi\sigma^2}}
\exp\!\left[-\frac{x^2}{2\sigma^2}\right],    
\end{equation}
 with zero mean and standard deviation $\sigma$. With this choice, under mean-field approximation, the value of  misinformation estimates to (see App.~\ref{app:A} for details) 
\begin{eqnarray}
I_{M} \;\approx\;
\frac{A}{2\sigma^2}
+
\frac{B}{4\sigma^4},
\label{eq:IMax_Gaussian}
\end{eqnarray}
where the mean-field coefficients $A$ and $B$ depend on the
distance statistics of the network and satisfy $0<A,B<1$. One can notice that the misinformation measure vanishes in the large-variance limit
as the extremal distortion is determined solely by the 
variance of the  Gaussian distribution. This means that when the intrinsic uncertainty of the global state dominates over network-induced
biases, the topology is irrelevant. When the true
distribution is broad and unstructured, finite perception errors
cannot be distinguished, and the network ceases
to act as an ``information-distorting medium''.}  As $\sigma \to 0$ in Eq.~\ref{eq:IMax_Gaussian}, this bound diverges, 
implying  that the Dirac-delta limit is ill-defined in this coarse-grained formalism 
and therefore requires  a separate treatment.

For non-Gaussian integrable and nonnegative PDFs, the estimate no longer reduces to a simple 
closed-form expression, but it can be evaluated, in principle, from 
Eq.~\ref{eq:continuous_I}. On the other hand, complete networks admit a closed-form solution due to their uniform connectivity. These two special cases---the Dirac delta distribution and the complete graph topology are examined in the following sections.}
\section{Complete Graph Topology}

The true distribution \(P(x)\) can be expressed in discrete form as~\cite{oppenheim_book}
\begin{equation}
    P(x) = \sum_{k} P(x_k)\,\delta\!\left(x - x_k\right),
\end{equation}
{\color{black} where the set $\{x_k\}$ denotes the discrete support points of the distribution, and the weights $P(x_k)$ satisfy the normalization condition $\sum_k P(x_k) = 1$.
}

In a complete graph, the shortest-path length between any two distinct
nodes is unity.  Hence, the error accumulated during the propagation of
information from node \(i\) to node \(j\) consists of a single Bernoulli
trial.  The perceived distribution \(Q^{(j)}(x)\) at any node \(j\) is
therefore
\begin{eqnarray}
Q^{(j)}(x) &=&
f\! \sum_{k} P(x_k)\,
\delta\!\left(x - x_k - \frac{1}{n-1}\right)
\nonumber\\
&&
+\, (1-f)\ \sum_{k} P(x_k)\,
\delta\!\left(x - x_k + \frac{1}{n-1}\right)
\nonumber\\
&&
+\, \frac{1}{n} \sum_{k} P(x_k)\,\delta\!\left(x - x_k\right),
\qquad \forall j.
\end{eqnarray}
{\color{black} Here, $f$ denotes the  fraction of the $(n-1)$ incoming states (from nodes other than $j$) that introduce a positive error. The first two terms represent the shifts $(x_k \to x_k + \tfrac{1}{n-1})$ and $(x_k \to x_k - \tfrac{1}{n-1})$ arising from the stochastic error process. The final term corresponds to the self-state of node $j$, which is perceived without distortion.
}

\medskip
\noindent
To analyse the behaviour of misinformation as the network size grows,
observe that the shift magnitude \(1/(n-1)\) vanishes in the limit
\(n\to\infty\).  Consequently, the shifted measures converge to the
original distribution \(P(x)\).  Moreover, by the law of large numbers,
the empirical fraction \(f\) converges to the Bernoulli bias parameter
\(r\).  
Thus, in the continuum limit the perceived distribution becomes
indistinguishable from the true distribution:
\begin{equation}
Q^{(j)}(x) \longrightarrow P(x), \qquad \text{as } n\to\infty.
\end{equation}

Because the KL divergence between identical distributions is zero
\cite{cover_2006, mackay_2003}, we obtain
\begin{equation}
    D_{KL}(P \Vert Q^{(j)}) \longrightarrow 0 
    \qquad \text{for all } j,
\end{equation}
which implies that the misinformation per node satisfies
\begin{equation}
    I_M \longrightarrow 0, 
    \qquad \text{as } n\to\infty.
\end{equation}

This behaviour is a direct consequence of the normalization scheme
introduced earlier, in which the maximum possible error is bounded by
unity.  
Because the error contribution per link decays as \(1/(n-1)\), a complete
graph becomes ``too well connected’’ for misinformation to accumulate in
large networks. Even a biased error process in this case cannot propagate over a
sufficiently long distance to produce a macroscopic distortion of the
global distribution.

\section{Analytical case study of a Dirac-Delta true state}
\label{sec:5}

{\color{black} In this section,  we begin with a general network} in which all nodes share an identical state, say \(a\).
In this case, the true-state probability density function reduces to a Dirac
distribution,
\begin{equation}
P(x) = \delta_{x,a}.
\end{equation}
For analytical convenience, we assume unbiased errors, \(r = \tfrac{1}{2}\), 
so that the probabilities of positive and negative errors are equal, as in 
Eq.~\ref{eq:binomial}. Under this protocol, node \(j\) can perceive the state 
of node \(i\) without distortion, but only when the shortest path between 
them has even length. Accordingly, the perceived distribution \(Q^{(j)}(x)\) 
decomposes as
\begin{eqnarray}
&&Q^{(j)}(x)
= Q^{(j)}(x = a) + Q^{(j)}(x \neq a)
\nonumber\\
&&= \frac{1}{n}\!\left[\,
    1
    + \sum_{m\in 2\mathbb{Z}_{\ge 0}}
        S\!\left(m,\tfrac{1}{2},\tfrac{m}{2}\right)
        b^{(j)}_m
    \right]\delta_{x,a}
    + Q^{(j)}(x \neq a), \nonumber \\
\end{eqnarray}
where \(b^{(j)}_m\) denotes the number of nodes at shortest-path distance
\(m\) from node \(j\). Evaluating the above expression at \(x=a\) yields
\begin{equation}
Q^{(j)}(a)
=
\frac{1}{n}\!\left[\,
    1
    + \sum_{m\in 2\mathbb{Z}_{\ge 0}}
        S\!\left(m,\tfrac{1}{2},\tfrac{m}{2}\right)
        b^{(j)}_m
\right].
\end{equation}

The KL divergence between the true and perceived distributions is then
\begin{equation}
D_{KL}\!\left(P \,\Vert\, Q^{(j)}\right)
    =
    \log n
    - \log\!\left[\,
        1 + \sum_{m\in 2\mathbb{Z}_{\ge 0}}
            S\!\left(m,\tfrac{1}{2},\tfrac{m}{2}\right)
            b^{(j)}_m
        \right].
\end{equation}
Averaging over all nodes gives the misinformation per node,
\begin{equation}
I_M
=
\log n
-
\frac{1}{n}\sum_{j=1}^{n}
\log\!\left[\,
    1 + \sum_{m\in 2\mathbb{Z}_{\ge 0}}
        S\!\left(m,\tfrac{1}{2},\tfrac{m}{2}\right)
        b^{(j)}_m
\right].
\label{eq:13}
\end{equation}

The second term in Eq.~\ref{eq:13} depends entirely on the network
topology and on the statistics of error propagation. Therefore, an
optimal topology—one that minimizes, maximizes, or stabilizes the
misinformation—must satisfy
\begin{equation}
\sum_{j=1}^{n}
\Delta\log\!\left[
    1 + \sum_{m\in 2\mathbb{Z}_{\ge 0}}
        S\!\left(m,\tfrac{1}{2},\tfrac{m}{2}\right)
        b^{(j)}_m
\right] = 0,
\label{eq:14}
\end{equation}
where $\Delta$ denotes variations induced by infinitesimal (or discrete)
changes in the shortest-path distribution $b^{(j)}_m$.  
Although Eq.~\ref{eq:14} is exact, closed-form optimality conditions arise 
only under additional structural assumptions.

For distance-regular graphs—whose geodesic-shell structure is fully
characterized by an intersection array~\cite{brouwer1989distance}—the
shell occupations $b^{(j)}_m$ are independent of \(j\), reducing
Eq.~\ref{eq:14} to
\begin{equation}
\sum_{m\in 2\mathbb{Z}_{\ge 0}}
    S\!\left(m,\tfrac{1}{2},\tfrac{m}{2}\right)\,
    \Delta b_m = 0,
\label{eq:dr}
\end{equation}
which depends solely on the coefficients of the intersection array. For asymptotically tree-like networks (e.g., Erd\H{o}s--R\'enyi graphs
\cite{bollobas1998random} or configuration models~\cite{newman2001random}),
$b_m$ admits the closed-form approximation $b_m \approx c^m$ for mean
branching factor $c$, leading to the analytic extremality condition
\begin{equation}
\sum_{m\in 2\mathbb{Z}_{\ge 0}}
    m\, S\!\left(m,\tfrac{1}{2},\tfrac{m}{2}\right)\,
    c^{\,m-1} = 0.
\label{eq:tree}
\end{equation}

More generally, since $b^{(j)}_m = (A^m \mathbf{1})_j$, the spectral
decomposition of the adjacency operator~\cite{chung1997spectral} yields
the closed-form spectral condition
\begin{equation}
\sum_{m\in 2\mathbb{Z}_{\ge 0}}
    S\!\left(m,\tfrac{1}{2},\tfrac{m}{2}\right)
    \sum_{k}
        m\,\lambda_k^{\,m-1}
        (\mathbf{v}_k^\top\mathbf{1})^2
    = 0,
\label{eq:spectral}
\end{equation}
where $\lambda_k$ and $\mathbf{v}_k$ are the eigenvalues and eigenvectors
of $A$. Thus, fully explicit optimality criteria arise whenever the
geodesic structure of the network can be written in closed form—either
through distance-regularity, asymptotic tree-likeness, or spectral symmetry.

\section{Numerical Results}

To complement the analytical results of Sec.~\ref{sec:5}, we performed numerical
experiments on three canonical network ensembles: Erd\H{o}s--R\'enyi (ER)
random graphs, Watts--Strogatz (WS) small-world networks, and
Barab\'asi--Albert (BA) scale-free networks.  
For each ensemble, we computed the misinformation $I_M$
defined in Eq.~\ref{eq:I_M}, averaging over \emph{twenty-five}
independent realizations.  
Unless stated otherwise, all simulations used $n = 5000$ nodes, and the
perception distributions $Q^{(j)}(x)$ were sampled with a bin width
$W \approx 2/(n-1)$ to resolve the error-induced deviations accurately. For finite networks, the perceived distributions \(Q^{(j)}(x)\) must be 
estimated using histograms. Because the maximum error accumulated along any 
path in a complete graph is $\pm 1/(n-1)$, the histogram bin width should 
be chosen to resolve this scale. In practice, a bin width of order $W \approx 2/(n-1)$
captures the full range of error-induced shifts without excessive smoothing. 
Larger bins mask small deviations and underestimate misinformation, whereas 
bins that are too fine lead to sparse counts and noisy estimates of 
\(Q^{(j)}(x)\).

{\color{black}To enable a viable comparison across networks with different structural properties, we normalize the misinformation by the maximum possible value achievable under the protocol. The normalized misinformation per node is therefore
\begin{equation}
I_M =
\frac{\frac{1}{n}\sum_{j=1}^{n} D_{KL}(P \Vert Q^{(j)})}
     {\mathcal{I}},
\label{eq:normalized_mi}
\end{equation}
where we choose $ \mathcal{I} = \log n$ for all the plots as this is a natural choice evident from Eq.~\ref{eq:13}.}

These three topologies were chosen because they span distinct geometric and 
dynamical regimes relevant to stochastic error propagation. 
Erd\H{o}s--R\'enyi graphs represent maximally random connectivity, with 
approximately Poissonian degree fluctuations~\cite{erdos_1959,haenggi_ieee_2009}. 
Watts--Strogatz networks interpolate between regular lattices and random 
graphs, capturing the coexistence of short paths and high clustering that 
shapes diffusion and communication processes~\cite{watts_nature_1998,newman_pnas_2001}. 
Barab\'asi--Albert networks produce heterogeneous, hub-dominated structures 
through preferential attachment, resulting in heavy-tailed degree 
distributions and nontrivial transport pathways~\cite{barabasi_science_1999,albert_nature_1999}. Because these ensembles differ sharply in geodesic organization, clustering, 
and degree heterogeneity, they induce markedly different patterns of error 
accumulation and dissipation. Together, they provide a representative 
cross-section of the topological regimes in which misinformation dynamics 
typically exhibit nonlinear amplification, suppression, or saturation 
behaviors.

\subsection{Erd\H{o}s--R\'enyi Networks}
For Erd\H{o}s--R\'enyi (ER) graphs $\mathcal{G}(n,p)$, we considered networks with 
$n = 5000$ nodes and varied the link probability~$p$, giving an average 
degree $\langle k \rangle = np$~\cite{erdos_1959,bollobas_2001}.  
Although few real-world systems follow a pure ER structure, the model is 
widely used as a null ensemble when links can be approximated as forming 
independently with equal probability~$p$%
\cite{erdos_1959,bollobas_2001}.  
Such approximations arise in relatively homogeneous settings, including 
classroom social networks~\cite{wasserman_1994}, local peer-to-peer 
wireless setups~\cite{li_wcmc_2003}, and laboratory collaboration 
groups~\cite{newman_pnas_2001}.  
Similarly, locally dense neuronal microcircuits exhibit partially random 
synaptic connectivity with experimentally observed connection probabilities 
of $5\%$--$30\%$%
\cite{song_plos_2005,lefort_neuron_2009,perin_pnas_2011}.  
For networks of size $n\sim 10^3$, these values correspond to link 
probabilities $p \approx 0.05$--$0.3$, yielding mean degrees in the range 
$\langle k \rangle \sim 50$--$300$, consistent with empirical densities.  
Although real systems also display clustering and spatial constraints, 
they are typically sparse with roughly probabilistic links, making the ER 
model a natural and widely used baseline for comparison.

\begin{figure}[t]
\centering
\includegraphics[width=0.48\textwidth]{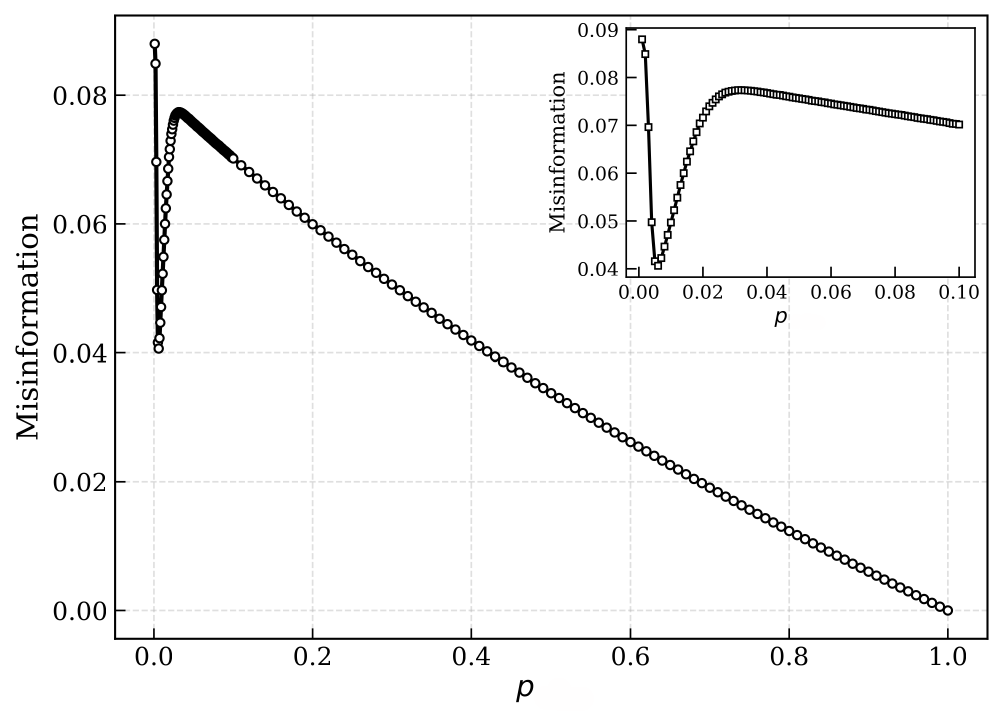}
\caption{\textbf{Normalized misinformation ($I_M$) as a function of the link 
probability ($p$) for an Erd\H{o}s--R\'enyi network with $n = 5000$ nodes 
and a delta-distributed true state.} The curve begins at a high 
misinformation level for the sparsest connected graphs and then drops 
rapidly to a shallow minimum ($I_M \approx 0.04$) immediately above the 
connectivity threshold, where the network remains almost tree-like and 
limits error accumulation. As $p$ increases further, $I_M$ rises again to a 
second, smaller peak around $p \approx 0.031$, reflecting the coexistence of 
long geodesic paths and emerging short cycles. Beyond this point, 
misinformation decreases monotonically, approaching zero as $p \to 1$, where 
uniformly short paths strongly suppress accumulated binomial errors. 
This overall pattern highlights a distinct double-peaked, nonmonotonic 
dependence of misinformation on network density.}
\label{fig:er_medium}
\end{figure}

Figure~\ref{fig:er_medium} reveals a striking feature of the 
Erd\H{o}s--Rényi network---the normalized misinformation exhibits a strongly 
non-monotonic dependence on the link probability, with two distinct local 
maxima appearing in the sparse regime. For the sparsest connected graphs, 
the misinformation begins at a high value and then drops rapidly to a shallow 
minimum just above the connectivity threshold, where the network remains 
almost tree-like and supports only a single long path between most node 
pairs, limiting the accumulation of errors. As the link probability 
increases further, additional edges shorten some geodesics while still 
preserving many long routes; this coexistence leads to a second, smaller 
maximum at slightly higher connectivity, reflecting amplification of 
path-accumulated binomial errors. Beyond this regime, the Erd\H{o}s--Rényi 
network becomes progressively more connected, and the average shortest path 
collapses toward its logarithmic limit~\cite{chung_lu2002}. This causes the 
misinformation to decrease monotonically and eventually approach zero when the network is fully connected. 
Thus, the largest distortions arise not at high connectivity but within a 
specific sparse range where information spreads widely yet still traverses 
error-prone long paths. This result is a structural feature emerging naturally from our 
model and, to our knowledge, not previously reported.
\begin{figure*}[t]
\centering
\includegraphics[width=0.87\textwidth]{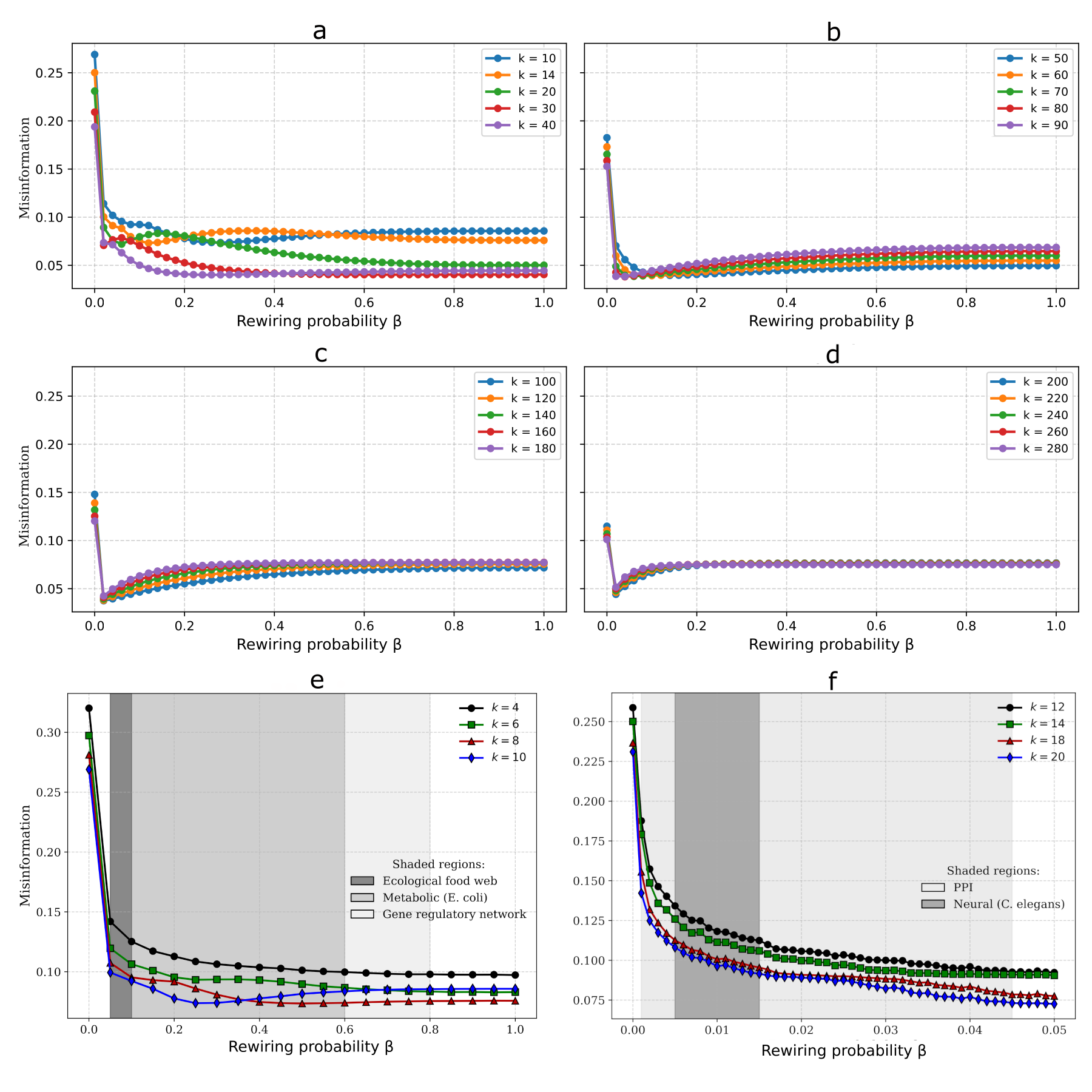}
\caption{\textbf{Normalized misinformation ($I_M$) as a function of the rewiring 
probability for Watts--Strogatz networks with $n = 5000$ nodes and a 
delta-distributed true state, shown for different average degrees (panels 
a–d).}  
The curves exhibit a clear non-monotonic trend: misinformation is highest 
in the highly ordered lattice (very low rewiring), decreases to a minimum 
at a small intermediate level of randomness, and then rises again before 
saturating in the fully randomized regime. This narrow small-world region 
balances local order with a few long-range shortcuts—enough to enable 
efficient communication while still permitting paths long enough for error 
accumulation—whereas both very ordered and highly randomized networks show 
elevated misinformation relative to this intermediate optimum.  
Panels (e) and (f) show the regimes occupied by real-world small-world 
networks and their corresponding misinformation levels, based on the 
effective rewiring estimates reported in 
Table~\ref{tab:biological_beta_two_col_final}.}
\label{fig:ws_medium}
\end{figure*}
\subsection{Watts--Strogatz Networks}
While Erd\H{o}s--R\'enyi graphs provide a useful baseline, their lack of 
clustering and structural heterogeneity limits their relevance for most 
real-world systems. Empirical studies across social, biological, and 
technological domains consistently show high clustering, modular 
organization, heterogeneous degree distributions, and short characteristic 
path lengths~\cite{watts_nature_1998,barabasi_science_1999,newman_siam_2003,
albert_rmp_2002,girvan_pnas_2002}. These features fall outside the ER 
framework and motivate the use of more structured generative models. The 
Watts--Strogatz (WS) model~\cite{watts_nature_1998} captures the emergence of 
small-world structure by randomly rewiring edges in a regular lattice with 
probability $\beta$. This parameter controls the balance between local 
clustering and global integration: $\beta = 0$ produces a highly clustered 
ring lattice with long path lengths, increasing $\beta$ introduces 
shortcuts that rapidly reduce these lengths yet largely preserving local 
neighborhoods, and $\beta = 1$ corresponds to a random graph with very low clustering.

Small-world patterns are widely observed across biological systems, although the underlying networks are not explicitly formed by rewiring. Neural 
connectomes, protein–interaction and gene-regulatory networks, metabolic 
pathways, and ecological food webs  exhibit high clustering paired with 
short paths~\cite{jeong_nature_2001,han2004evidence,albert_jcell_2005,
ravasz_science_2002,barabasi2004network,montoya_jtb_2002,amaral_pnas_2000}. 
Comparisons with WS reference networks suggest that many biological systems 
occupy an intermediate effective rewiring regime, typically 
$\beta \approx 0.01$--$0.2$~\cite{latora2001efficient,sporns2004small,
humphries2008network,kaiser2004spatial}. Table~\ref{tab:biological_beta_two_col_final} 
lists effective rewiring values inferred from such empirical comparisons.

Figures~\ref{fig:ws_medium} show that misinformation depends jointly on the 
number of nearest neighbors and the amount of rewiring. A small amount of 
randomness produces the lowest misinformation, defining a narrow 
small-world regime in which long-range shortcuts substantially reduce path 
lengths while local clustering remains intact. Very low rewiring yields a 
highly ordered lattice where information spreads diffusively and errors 
accumulate along long routes. Conversely, excessive rewiring erodes the 
intermediate-scale organization that buffers local fluctuations, causing 
the network to behave more like a random graph with noise-amplifying 
interactions. This competition between structural order and topological 
randomness explains the observed non-monotonic dependence of misinformation 
on $\beta$.

Within this transition, an intermediate rewiring level forms an effective 
critical regime: local neighborhoods remain coherent, while a modest number 
of shortcuts provide efficient long-range coordination and suppress error 
accumulation~\cite{watts_nature_1998,strogatz2001}. This crossover from 
diffusion-dominated to shortcut-mediated transport mirrors phenomena in 
disordered media and complex networks~\cite{barrat2008,dorogovtsev2008}. 
Consequently, the minimum misinformation arises from the interplay between 
local structure (set by nearest neighbors) and the partial randomness 
introduced by rewiring, a balance that supports robust and efficient 
information flow in many natural and engineered systems.

\begin{table*}[t]
\centering
\small
\begin{tabular}{|p{3.1cm}|p{1.6cm}|p{1.8cm}|p{1.4cm}|p{1.9cm}|p{6.7cm}|}
\hline
\textbf{Biological network} 
& \textbf{Size $N$} 
& \textbf{Half-degree $k/2$} 
& \textbf{Path length $L$} 
& \textbf{Effective $\beta$} 
& \textbf{Interpretation aligned with our results} \\[3pt]
\hline

Neural (\emph{C.~elegans}) 
& $\sim 3\times10^2$ 
& 5--7 
& 2.5--3.0 
& 0.005--0.015 
& Strongly ordered regime with a steep drop in misinformation even at very low~$\beta$, matching the fast-decay behavior predicted by our ordered-network theory. \\[12pt]
\hline

Metabolic (\emph{E.~coli}, human) 
& $\sim 10^3$ 
& 1--3 
& 2.9--3.2 
& 0.05--0.6 
& Almost-ordered to random regime; misinformation shows a slow, nearly monotonic decrease—consistent with the gradual, hub-moderated suppression predicted for low-degree biological networks. \\[12pt]
\hline

Gene regulatory networks 
& $10^3$--$10^4$ 
& 1--3 
& 3--4 
& 0.04--0.8 
& Traverses ordered to random regimes depending on organism and condition; misinformation decays weakly due to sparse connectivity and localized regulatory motifs. \\[12pt]
\hline

Protein–protein interaction (PPI) 
& $10^3$--$10^5$ 
& 4--10 
& 3--5 
& 0.001--0.045 
& Deeply ordered regime dominated by hub-centric architectures; predicts a sharp decline in misinformation with increasing~$\beta$, consistent with strong hub buffering seen in large PPIs. \\[12pt]
\hline

Ecological / food-web networks 
& $10^2$--$10^3$ 
& $\sim$2--3 
& 2.5--4 
& 0.02--0.1 
& Almost-ordered to moderately random regime; misinformation remains nearly constant with a shallow decline, reflecting low connectivity and trophic modularity. \\[10pt]
\hline
\end{tabular}

\caption{%
Approximate effective rewiring probabilities ($\beta$) for representative biological networks, obtained by matching empirical characteristic path lengths ($L$) to Watts--Strogatz predictions~\cite{newman_pre_1999,newman_prl_2000}. 
The resulting interpretations show how biological systems occupy ordered, intermediate, or random topological regimes—mirroring the misinformation behaviors observed in our simulations.}
\label{tab:biological_beta_two_col_final}
\end{table*}

\begin{figure}[h]
    \centering
    \includegraphics[width=0.47\textwidth]{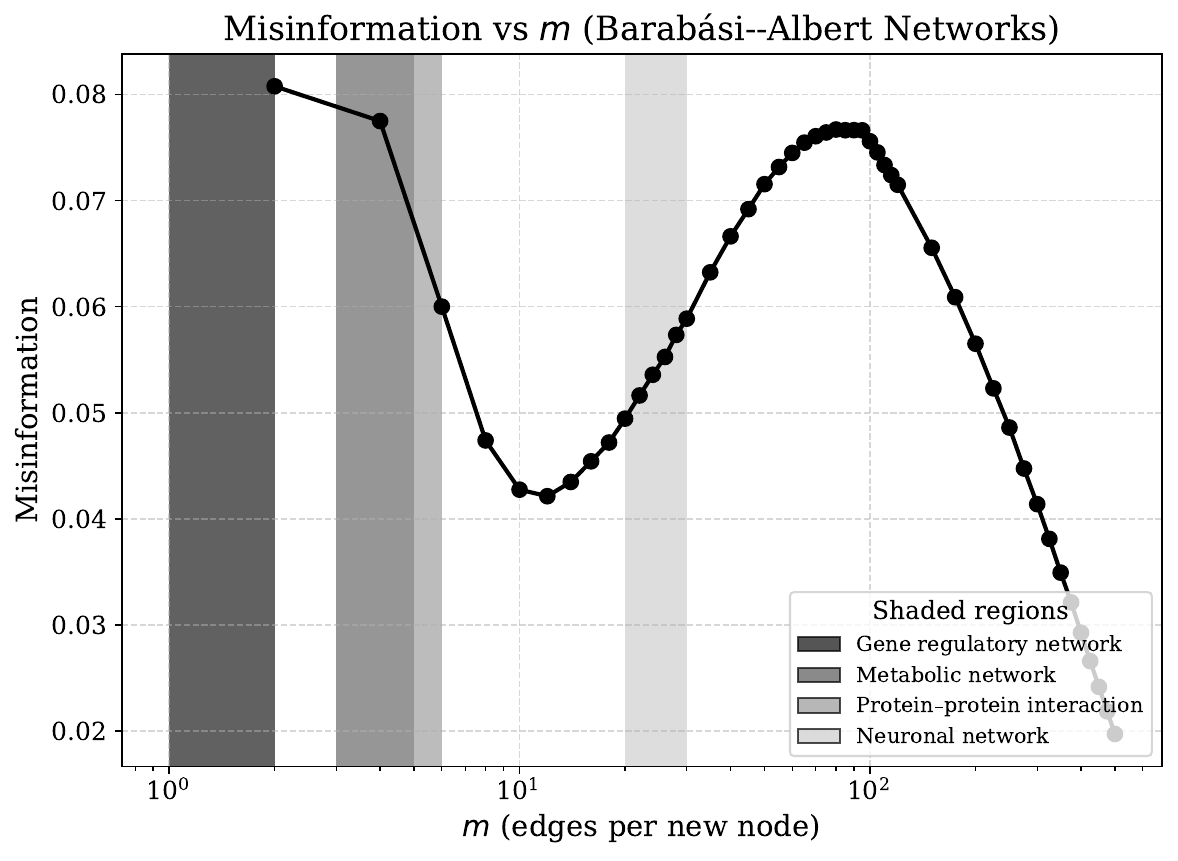}
    \caption{\textbf{Normalized misinformation ($I_M$) versus the number of edges added per 
new node ($m$) in Barab\'asi--Albert networks with $n = 5000$ nodes and a 
$\delta$-distributed true state.}  
Misinformation initially decreases monotonically with increasing $m$, 
reflecting the well-known role of hubs in enhancing robustness and 
stabilizing information flow in scale-free structures~\cite{barabasi_science_1999,
albert2000error,barrat2008}.  
Beyond an intermediate threshold, however, misinformation increases again, 
indicating a regime in which hub-driven fluctuations accumulate and propagate 
through the network.  
In the extremely sparse limit (small $m$), BA networks outperform random and 
small-world networks of comparable density, owing to their highly connected 
hubs, which act as efficient integrators and provide redundant communication 
pathways. Shaded regions show the regimes occupied by real-world scale-free 
networks and their corresponding misinformation levels, based on the comparable connectivity estimates reported in 
Table~\ref{tab:bio_SF_final_expanded}.
}
    \label{fig:BA}
\end{figure}

\begin{table*}[t!]
\centering
\small
\begin{tabular}{|p{3.4cm}|p{3.2cm}|p{7.1cm}|c|c|}
\hline
\textbf{Biological system} 
& \textbf{Example hubs} 
& \textbf{Functional role} 
& $\boldsymbol{\gamma}$ 
& \textbf{BA $m$} \\[4pt]
\hline

Metabolic network (\emph{E. coli}) 
& ATP, NADH, H$_2$O, CO$_2$ 
& Hub metabolites act as universal biochemical carriers that integrate 
  multiple pathways, stabilize global flux balance, and buffer stochastic 
  fluctuations arising from local reaction noise~\cite{jeong2000}. 
& 2.2--2.4 
& 3--5 \\[14pt]
\hline

Gene-regulatory network 
& p53, NF-$\kappa$B 
& Master transcription factors regulate many downstream targets, enforce 
  coordinated responses, maintain expression stability, and reduce error 
  propagation in regulatory cascades~\cite{wagner2001small}. 
& 2.1--2.5 
& 1--2 \\[14pt]
\hline

Protein--protein interaction (PPI) network 
& Kinases, chaperones 
& Highly connected proteins integrate signaling modules, buffer proteomic 
  perturbations, stabilize folding environments, and prevent 
  miscommunication within interaction pathways~\cite{jeong_nature_2001}. 
& 2.4--2.7 
& 3--6 \\[14pt]
\hline

Neuronal connectome 
& Thalamus, precuneus, rich-club hubs 
& High-degree cortical hubs support long-range integration, enable rapid 
  routing between distributed brain modules, and mitigate noise 
  accumulation in multi-step communication processes~\cite{eguiluz2005scale}. 
& 2.0--2.3 
& 20--40 \\
\hline
\end{tabular}

\caption{Representative biological scale-free networks with example hubs, functional roles relevant to misinformation suppression, and their characteristic 
degree-distribution exponents $\gamma$ and approximate Barabási--Albert 
parameter $m$ ($\langle k \rangle \approx 2m$). Neuronal ranges vary due to 
differences in connectome resolution, parcellation, and thresholding across 
datasets~\cite{bullmore2009,hagmann2008mapping,modha2010network,van2010exploring}.
}
\label{tab:bio_SF_final_expanded}
\end{table*}
\subsection{Barab\'asi--Albert Networks}

Scale-free networks generated by the Barab\'asi--Albert (BA) preferential 
attachment model~\cite{barabasi_science_1999} provide a natural framework 
for examining the role of heterogeneous connectivity in misinformation. In 
this model, each newly added node attaches to $m$ existing nodes with 
probability proportional to their degrees, producing a power-law degree 
distribution
\begin{equation}
P(k) \sim k^{-\gamma}, \qquad \gamma \approx 3,
\end{equation}
and yielding a small number of highly connected hubs alongside many 
low-degree nodes. Because each new node introduces exactly $m$ edges, the 
total number of edges grows linearly with network size, giving an exact 
average degree $\langle k \rangle = 2m$. This heavy-tailed structure 
captures an organizational pattern observed broadly in biological and 
technological systems.

The biological relevance of scale-free architecture is well documented. 
Metabolic networks~\cite{jeong2000,wagner2001small}, gene-regulatory 
circuits~\cite{lee2002transcriptional,babu2004structure}, and protein–protein 
interaction networks~\cite{jeong_nature_2001,han2004evidence} all display 
degree exponents between 2 and 3. In such systems, hubs frequently 
correspond to essential components—such as ATP in metabolism or master 
transcriptional regulators—that coordinate activity across multiple 
pathways and support robust communication. For context, 
Table~\ref{tab:bio_SF_final_expanded} compares several biological 
scale-free networks with a BA network of matched average degree. Although 
empirical exponents vary across the range 2–3, the BA model consistently 
produces an exponent close to~3.

Turning to our numerical results, Fig.~\ref{fig:BA} shows that 
misinformation in BA networks depends non-monotonically on the attachment 
parameter~$m$. For small values of $m$, misinformation decreases rapidly: 
each additional edge enhances long-range communication and reduces the 
depth of error-prone paths. Beyond an intermediate connectivity level, 
misinformation begins to rise again and eventually saturates, reflecting the 
interplay between shortened geodesics and the heterogeneous load 
concentrated on hubs.
\begin{figure*}[t]
    \centering
    \includegraphics[width=0.80\textwidth]{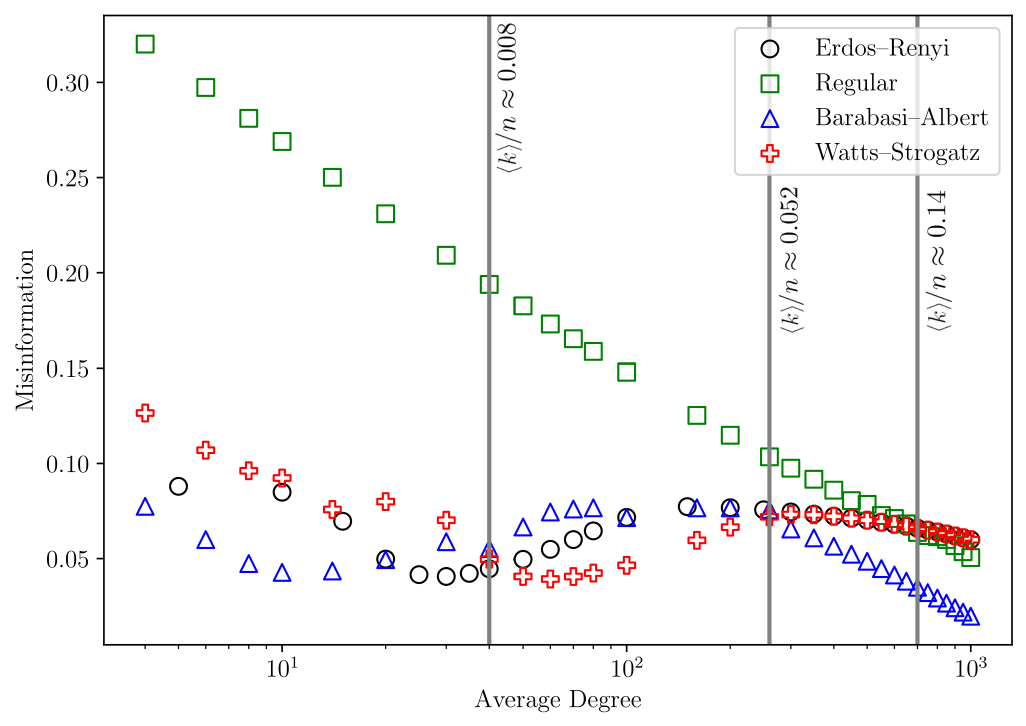}
    \caption{\textbf{A comparison of misinformation across regular, ER, WS, and BA networks, 
all constructed with $n = 5000$ nodes.}  
{\color{black} Regular networks consistently exhibit the highest misinformation in the sparse to intermediate-dense regimes due to 
their long path lengths and absence of shortcuts.  
In the extremely sparse regime, ER and BA networks behave similarly, with BA 
showing slightly lower misinformation because early-formed hubs provide 
local aggregation.   As connectivity increases into an intermediate (but still relatively low)
range, a clear crossover emerges: WS networks achieve the lowest
misinformation as they enter the small-world regime, where the introduction
of a modest number of shortcuts sharply reduces characteristic path lengths
while largely preserving local structure. In the intermediate-dense to
dense-connected regime, BA networks outperform all other topologies, yielding
the smallest misinformation among the classes considered.
The WS curve shown here corresponds to a rewiring probability of $\beta = 0.1$, which lies within the small-world regime. The black solid lines represent the transition points, where $\langle k \rangle$ denotes the average degree.}}
    \label{fig:comparison}
\end{figure*}
\subsection{Hierarchy of network topologies under  error-propagation}

{\color{black}
A notable outcome of our analysis is that the relative performance of ER,
WS, and BA networks varies with sparsity, revealing \emph{four distinct
regimes} (see Fig.~\ref{fig:comparison}):

(a) In the \emph{extremely sparse regime} ($ \langle k \rangle /n \lesssim 0.008$)---where the
network is barely connected, our results suggest that BA networks tend to
exhibit the lowest misinformation among the four canonical topologies.
In this limit, ER and BA behave similarly: both reduce to thin branching
structures in which misinformation grows slowly with connectivity, while the
presence of a few BA hubs may provide additional local aggregation that
slightly suppresses fluctuations. WS networks, by contrast, remain dominated
by regular-lattice behavior at very low rewiring and thus can suffer from
long path lengths that accumulate more error.

(b) As connectivity increases into the \emph{low to intermediate regime}
($0.008 \lesssim  \langle k \rangle /n \lesssim 0.052$), a surprising crossover appears:
WS networks begin to outperform BA networks and achieve the minimum
misinformation. WS graphs enter the small-world regime early; moderate
increases in degree introduce many shortcuts without yet producing the
highly heterogeneous hub structure characteristic of BA networks. This
combination of short path lengths and relatively uniform degrees appears to delay the
onset of error accumulation. 


(c) In the regime of intermediate–high connectivity
($0.052 \lesssim  \langle k \rangle/n \lesssim 0.14$), the ordering becomes such
that BA networks again tend to exhibit the lowest misinformation and appear
to be the most robust topology. In this regime,
moderate increases in connectivity shorten many geodesics in ER and WS
graphs, but the heterogeneous degree distribution of BA networks
concentrates information flow through a relatively small set of high-degree
nodes. This may help to limit the uncontrolled spread of independent
pathwise errors. Regular lattices are the worst because their long geodesic
distances continue to accumulate errors.

(d) In the comparatively denser limit ($ \langle k \rangle /n \gtrsim 0.14$), the hierarchy
again places BA networks in the most robust position. In this regime,
regular graphs also gain substantial robustness. As their connectivity
increases, their highly symmetric structure begins to suppress errors
more effectively. This allows them to outperform both ER and WS networks. BA
networks, however, retain their apparent advantage and yield the smallest
misinformation among the four classes.
}

\section{Effect of Biased Pathwise Errors}
In our model, the information that node $j$ receives about node $i$
accumulates stochastic errors along the shortest path of length $d_{ij}$.
If $u$ of the $d_{ij}$ steps contribute positive increments and the rest
contribute negative ones, the perceived state takes the form given in
Eq.~\ref{eq:perceived}, where $u$ follows a binomial distribution with
parameters $(d_{ij}, r)$.  
The parameter $r$ controls a directional bias: for $r = 1/2$ the noise is
symmetric, whereas $r \neq 1/2$ introduces a preference for either positive
or negative updates.  
One might expect such a bias to modify the perceived distribution by
introducing skewness or broadening.  
However, the actual effect is far simpler: the bias induces only a uniform
shift of the distribution, while its overall fluctuation structure remains
unchanged.

\subsection{Mean and fluctuation structure of the accumulated error}

To make the impact of the bias explicit, we decompose the binomial variable 
into its mean and zero-mean fluctuation:
\begin{equation}
    u = d_{ij} r + \delta u,
    \qquad  
    \overline{\delta u} = 0,
    \qquad 
    \mathrm{Var}(\delta u) = d_{ij} r (1-r).
    \label{eq:binomial_decomposition}
\end{equation}
This separates the two essential components of the binomial error process.  
The term $d_{ij} r$ represents the expected number of positive increments 
along the path of length $d_{ij}$, whereas the deviation 
$\delta u = u - d_{ij} r$ captures zero-mean fluctuations around this drift.  
Because $u \sim \mathrm{Bin}(d_{ij}, r)$, these fluctuations have variance 
$d_{ij} r (1-r)$ and quantify the intrinsic randomness of pathwise error 
accumulation.  
This decomposition makes clear how the bias parameter influences 
information transfer: the mean shifts with $r$, while the fluctuation 
structure remains unchanged.

Substituting~\eqref{eq:binomial_decomposition} into~\eqref{eq:perceived} 
yields the exact expression
\begin{align}
    X^{(j\leftarrow i)}
    &=
      X_i 
    + \frac{1}{n-1} \left[ 2(d_{ij} r + \delta u) - d_{ij} \right] \nonumber\\
    &=
      X_i
    + \frac{d_{ij}(2r - 1)}{n-1}
    + \frac{2}{n-1}\,\delta u.
    \label{eq:drift_form}
\end{align}

Equation~\eqref{eq:drift_form} makes the structure of the propagated error 
explicit. The first term is a deterministic drift, proportional to $(2r - 1)$ that grows linearly with the shortest-path length $d_{ij}$, and vanishes in the 
unbiased case $r = 1/2$.  
The second term contains the fluctuations, proportional to $\delta u$, 
which remain symmetric for all values of $r$.  
Thus, the bias affects only the mean shift and leaves the fluctuation 
distribution---its symmetry, tails, and overall shape-unaltered, apart from 
the standard binomial variance $d_{ij} r (1-r)$, which is symmetric under 
$r \leftrightarrow (1-r)$.

\subsection{Translation of the perceived distribution}
Let $Q^{(j)}_0(x)$ denote the distribution of $X^{(j \leftarrow i)}$ in the 
unbiased case $r = 1/2$.  
From Eq.~\eqref{eq:drift_form}, the perceived value under general $r$ can be 
written as
\begin{equation}
    X^{(j \leftarrow i)}(r)
    =
    X^{(j \leftarrow i)}(1/2)
    +
    \Delta\mu_{ij},
\end{equation}
where the shift is
\begin{equation}
    \Delta\mu_{ij}
    =
    \frac{d_{ij}(2r - 1)}{n - 1}.
\end{equation}
The fluctuation term $2\delta u/(n-1)$ does not appear explicitly because its 
distribution is independent of $r$: the bias modifies only the mean of 
$X^{(j \leftarrow i)}$, while the centered binomial fluctuations remain 
unchanged.  
Thus, the effect of bias is a pure translation,
\begin{equation}
    Q^{(j)}_r(x)
    =
    Q^{(j)}_0\!\left(x - \Delta\mu_{ij}\right),
    \label{eq:shift_relation}
\end{equation}
meaning that each perceived value is shifted by a constant amount.

To understand why Eq.~\eqref{eq:shift_relation} preserves the functional 
shape of the distribution, consider the case where the true state $X_i$ is 
Gaussian.  
The perceived value $X^{(j \leftarrow i)}$ is the sum of this Gaussian 
variable, a constant drift, and a centered binomial fluctuation.  
Because the fluctuation term is a sum of independent Bernoulli increments, 
it is approximately Gaussian by the central limit theorem~\cite{feller_prob}; 
and the sum of two Gaussian variables is Gaussian~\cite{papoulis}.  
Hence the overall shape is preserved.  
This drift–fluctuation decomposition mirrors classical treatments of 
stochastic processes~\cite{vankampen_stochastic}. Equation~\eqref{eq:shift_relation} therefore shows that \emph{bias does not 
deform the distribution}: it introduces no broadening, skewness, or change in 
shape.  
It generates only a uniform drift proportional to the geodesic distance and 
the imbalance $(2r - 1)$, while the statistical structure of the underlying 
fluctuations remains invariant.

 \begin{figure*}[t]
\includegraphics[width=0.90\textwidth]{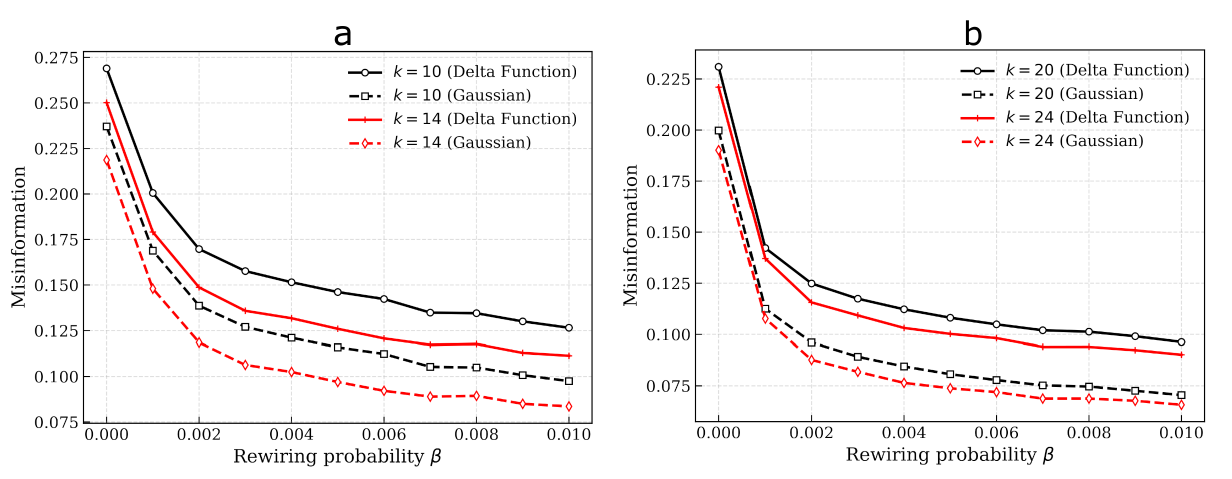}
\caption{\textbf{Comparison of misinformation for $\delta$-distributed and narrow Gaussian initial states as a function of the rewiring probability $\beta$.}
The Gaussian distribution is centred at the same value as the $\delta$ distribution and has a small standard deviation of $10^{-4}$.
The result is independent of the mean of the Gaussian distribution.
Even this extremely small intrinsic variance reduces the peak misinformation compared to the $\delta$ case, indicating that broader true states partially absorb accumulated binomial fluctuations.
}
    \label{fig:combined}
 \end{figure*}
\subsection{Consequences for misinformation}

The KL divergence quantifying misinformation at node $j$ becomes
\begin{equation}
    D_{KL}\!\left(
        P(x)
        \,\Vert\,
        Q^{(j)}_r(x)
    \right)
    =
    D_{KL}\!\left(
        P(x)
        \,\Vert\,
        Q^{(j)}_0(x - \Delta\mu_{ij})
    \right),
\end{equation}
so the only influence of $r$ enters through the translation $\Delta\mu_{ij}$.  
When the true distribution $P(x)$ is Gaussian, this dependence reduces to a 
simple quadratic function of the drift, since a translation of a Gaussian 
changes only its mean.  
Thus, varying the bias probability cannot generate any new qualitative behaviour in the 
misinformation. It shifts all perceived distributions coherently while 
the shapes and dispersions remain unchanged.  
This decomposition into a deterministic drift and a shape-preserving 
fluctuation term mirrors classical results for biased random walks and 
additive binomial noise~\cite{feller_prob,vankampen_stochastic,cover_2006}.  
Similarly, in our network setting,  the bias induces only a 
uniform shift, without altering the form of the perceived 
distribution.

In the limit where the true distribution collapses to a single discrete 
value, the impact of non-symmetric biases ($r \neq 1/2$) remains finite.  
Although the drift scales as $(2r - 1)d_{ij}/(n-1)$, the binomial error 
mechanism ensures that $Q^{(j)}(x)$ continues to assign nonzero probability 
to the true state, preserving overlap between the distributions.  
This prevents divergences that would otherwise arise in the continuous KL 
divergence between shifted delta functions.  
Moreover, the normalized misinformation stays bounded because both the 
observed and maximum possible misinformation are evaluated over the same 
finite discrete state space.  
Hence, in the $\delta$-limit, the effect of bias is simply a smooth and 
finite shift in the perceived distribution.

A key physical implication is that biased errors do not increase the 
uncertainty of the information flowing through the network.  
Instead, they produce a coherent, topology-dependent displacement of the 
perceived state.  
Consequently, systematic misinformation accumulates along longer paths, 
making agents consistently incorrect rather than increasingly uncertain.  
This mechanism aligns with phenomena observed in social systems, where 
groups may converge to shared but biased narratives while maintaining 
internal confidence~\cite{delvicario2016,sunstein2009}.  
Similar effects appear in technological and biological settings---in 
distributed sensing, small measurement biases can shift global consensus 
without increasing estimated noise~\cite{olfatisaber2007}, and in 
regulatory or signaling networks, biased transmission can lead to coherent 
misregulation rather than merely noisier responses.

\section{Discussion \& Conclusions}

Our results show that misinformation in complex large networks is shaped by an 
interplay between geodesic structure, stochastic error accumulation, and 
topological organization. Rather than depending solely on density or degree, 
misinformation emerges from how network geometry transforms local 
binomial errors into global distortions. This perspective unifies the 
behaviour of Erd\H{o}s--R\'enyi (ER), Watts--Strogatz (WS), and 
Barab\'asi--Albert (BA) networks within a single analytical and numerical 
framework.

In ER networks, misinformation displays a distinctive double-peaked profile 
as the link probability varies. Near the percolation threshold, long 
geodesic chains amplify accumulated fluctuations, whereas at slightly higher 
densities the coexistence of extended paths and emerging cycles produces a 
second, weaker maximum~\cite{newman_2010,barrat2008}. Only at high densities—where 
shortest paths collapse logarithmically—does misinformation sharply 
decrease. This behaviour highlights a general principle: misinformation is 
maximized when the network is sufficiently connected to propagate signals 
widely yet sparse enough to preserve long, error-amplifying routes.

WS networks reveal a complementary trend. A small amount of rewiring creates 
a narrow small-world regime in which misinformation is minimized. Sparse 
random shortcuts reduce path lengths dramatically while retaining strong 
local structure~\cite{watts_nature_1998,strogatz2001}. This structural 
crossover resembles transport optimisation in disordered media 
\cite{dorogovtsev2008}. Notably, empirical biological networks—neural, 
metabolic, genetic, and ecological—often fall precisely within this 
low-to-intermediate randomness range~\cite{sporns2004small,jeong2000,alon2007introduction}, 
aligning with  observation that small-world architectures suppress 
global distortion.

{\color{black}BA networks exhibit the lowest level of misinformation  both in the sparse and intermediate-density connectivity regimes.} 
High-degree hubs act as robust integrators that shorten effective distances, 
dilute noise, and stabilise information flow~\cite{barabasi_science_1999,albert2000error,jeong_nature_2001,han2004evidence}. 
This result provides a quantitative explanation for the resilience of 
scale-free systems---heterogeneity and hub dominance inherently suppress the 
accumulation of pathwise errors. {\color{black} Although BA networks are known to facilitate rapid propagation of both
information and errors due to their hub-dominated architecture~\cite{barthelemy_prl_2004,wang_rsos_2018}, our results
show that, within a static error-accumulation model, they exhibit robustness to path induced errors. This apparent contrast
between quick dynamical error propagation and suppressed misinformation is a
 nontrivial feature of scale-free topology. This feature highlights an
interesting phenomenon that merits investigation in future
work.}

A central analytical contribution of this work is the demonstration that 
directional bias in the error process creates only a uniform drift in the 
perceived state, without altering the fluctuation structure. The biased 
process splits into an additive drift term and a shape-preserving 
noise term, similar to classical stochastic-process theory 
\cite{feller_prob,vankampen_stochastic}. Consequently, biased errors produce 
\emph{coherent misinformation} rather than uncertainty. Our formalism connotes several
observed phenomena in social systems, where groups converge to internally 
confident but biased narratives~\cite{sunstein2009,delvicario2016}—and in 
technological or biological networks, where small sensor biases 
imbalances drive global misregulation~\cite{olfatisaber2007}.

Comparing delta-distributed and narrow Gaussian true states reveals 
that even minimal intrinsic stochasticity reduces achievable misinformation 
(consistent with Eq.~\ref{eq:IMax_Gaussian}) and smooths its dependence on 
connectivity (Fig.~\ref{fig:combined}). We show that an intrinsic heterogeneity in the true global state can act as a buffer to error propagation as small errors remain partly absorbed within the 
spread of the true state, whereas a perfectly sharp state is maximally 
fragile.

Overall, our findings reveal a unified physical picture: 
\emph{misinformation is governed by the geometry of geodesic paths and the 
structure of fluctuations, and different network topologies impose distinct 
transformation rules on how local stochastic errors accumulate}. Scale-free 
networks suppress distortion through hub-dominated integration; small-world 
networks exploit a balance between clustering and global shortcuts; and ER 
networks amplify noise near structural transitions.

\textcolor{black}{The framework introduced in our work is deliberately static, or asymptotic in time: information is assumed to propagate along shortest paths in a single pass, without any time dependence, feedback loops, memory effects, or multi-round updating. This abstraction deviates from many real-world misinformation propagation processes, where delays, competition between narratives, repeated exposures, and corrective mechanisms play central roles~\cite{centola_science_2010,vicario_2016,vosoughi_science_2018,newman2018}. Our objective, however, is not to reproduce the full temporal complexity of social or biological communication, but rather to quantify the contribution of network topology to error propagation under minimal and well-controlled dynamical assumptions. In this sense, shortest-path propagation provides a natural assumption that captures the most direct routes of error propagation without losing the  interpretation of the role of connectivity. More general schemes, such as diffusion-based spreading, averaging over multiple paths, or incorporating adaptive, or feedback-driven dynamics, would  enrich the model, but at the cost of increased complexity. We therefore view the present framework as a foundational baseline rather than a comprehensive analysis of the error spreading process. We emphasize that our purpose is to establish an  analytically tractable limit on misinformation. Such limits can serve as a reference point for future realistic extensions incorporating temporal evolution, alternative routing mechanisms, and heterogeneous agent behavior.}

\textcolor{black}{Our model assumes that each transmission introduces an independent error drawn from a common distribution. We adopt this assumption to ensure analytical tractability and to isolate the influence of network topology on error propagation. Within this assumption, our framework already accommodates certain forms of heterogeneity, including asymmetric binomial noise, and Gaussian initial state distributions, thereby demonstrating that the qualitative behavior of error propagation is robust to changes in the distribution-shape, variance, and skewness. Nevertheless, the present framework does not capture correlated noise, either temporal or spatial, and  source-dependent biases, such as misinformation introduced by some influential nodes, coordinated media outlets, or tightly coupled communities~\cite{mezard_1986,lazer_science_2018,delvicario2016}. Allowing node-specific error distributions, correlated transmissions, or adversarial sources could give rise to qualitatively new phenomena, including localization of error, amplification by central actors, or nontrivial spatial clustering. Exploring such effects would require relaxing the assumption of independence and extending the framework to incorporate heterogeneous, correlated, or memory-dependent noise processes~\cite{gleeson_prx_2016,pastor_rmp_2015}. We leave these directions for future work, while we emphasize that the present framework establishes a baseline upon which more elaborate models can be built.}

\textcolor{black}{Since the KL divergence measures the mismatch between a true distribution and its distorted form, it turns out to be the simplest yet standard choice in our framework to measure the misinformation~\cite{kullback_1951,cover_2006}. In principle, one can quantify misinformation between a perceived distribution and the true distribution using a more general class of asymmetric divergence measures, namely the R\'enyi divergences.  The R\'enyi divergences
form a one-parameter family of divergence measures indexed by an order 
parameter $\alpha>0$, $\alpha \ne 1$.  The KL divergence appears as a special (limiting) case of 
the R\'enyi divergence in the limit $\alpha \to 1$~\cite{renyi_1961}.  More generally, R\'enyi 
divergences of different orders emphasize different regions of the 
distributions: values $\alpha<1$ give more weight to rare events and 
distribution tails, whereas values $\alpha>1$ place greater weight on 
high-probability events~\cite{cover_2006,erven_ieee_2014}. However, as shown in App.~\ref{app:B}, in certain limits the R\'enyi divergences are related to the KL divergence in a simple manner. Specifically, (i) when the true-state probability density function is a Dirac-delta distribution, the R\'enyi divergences reduce to the KL divergence, irrespective of the value of $\alpha$, implying the misinformation remains unchanged; (ii) when the true distribution is sufficiently broad, the R\'enyi divergences become proportional to the KL divergence, with proportionality factor $\alpha$, implying that the misinformation is trivially  scaled. Thus in both these cases, our main results including the relative ranking of network topologies and the presence of crossover behavior remain unchanged. For more general cases, the impact of $\alpha$ on the main qualitative trends, such as the relative ranking of network topologies and the presence of crossover behavior, remains an interesting question and is left for future investigation.}

Overall, our theoretical and numerical results lay a 
foundation for a statistical model of information distortion in complex 
networks, providing both insight and a versatile analytical 
tool for future studies. In addition to the directions mentioned above, extending the model 
to weighted, directed, or multilayer networks would allow the 
analysis of systems with richer dynamical features. Incorporating 
correlated or memory-dependent errors could reveal how temporal persistence 
modifies misinformation propagation. From an optimization perspective, one 
could also search for cost-constrained rewiring strategies that will minimize global 
distortion or identify networks that maximize the robustness under 
biased noise. Finally, coupling this framework with 
agent-based dynamics or adaptive topology could shed light on how real 
systems evolve to balance efficiency, robustness, and misinformation 
suppression.

\section*{Author Contributions}

\textbf{Archan Mukhopadhyay}: Conceptualization (lead), Fomal Analysis (equal),  Methodology (equal), Supervision (lead), Writing- Original draft (lead). 
\textbf{Jens Christian Claussen}: Conceptualization (supporting), Supervision (supporting), Validation (equal).
\textbf{Saikat Sur}: Formal Analysis (equal), Methodology (equal), Supervision (supporting), Writing- Reviewing and Editing (lead), Visualization (supporting). \textbf{Rohitashwa Chattopadhyay}: Validation (equal), Visualization (lead), Data curation (lead), Writing- Reviewing and Editing (supporting).

\section*{Acknowledgements}
SS acknowledges the support of the Department of Chemical and Biological Physics and AMOS at the Weizmann Institute of Science, Israel, where a significant part of this work was carried out. JCC acknowledges financial support from the EPSRC through grant EP/V048740/2. AM acknowledges support from the School of Computer Science, University of Birmingham, and the Centre for Ecological Sciences, Indian Institute of Science, Bengaluru, where parts of this work were conducted. His stay at the University of Birmingham was supported by the EPSRC through grant EP/V048740/2, and his stay at the Indian Institute of Science, Bengaluru, was supported by the SERB (DST, Government of India) through project no. PDF/2023/001151. AM is currently employed at M. S. Ramaiah University of Applied Sciences. RC gratefully acknowledges the generous allocation of computing resources by the Department of Theoretical Physics (DTP), Tata Institute of Fundamental Research (TIFR), along with related technical assistance from Kapil Ghadiali and Ajay Salve.

\section*{Data Availability Statement}
The data supporting the findings of this study are available from the corresponding author upon reasonable request.

\section*{Conflicts of Interest}
The authors have no conflicts to disclose.



\section*{References}

\bibliography{manuscript}


\appendix
\section{Mean-field estimate of misinformation for a Gaussian true distribution}\label{app:A}

Let $P(x)$ denote a normalized  probability density function and $Q(x)$ be another probability density that
differs from $P(x)$ by a small perturbation $\Delta(x)$, i.e.,
\begin{equation}
Q(x) = P(x) + \Delta(x),
\label{eq:Q_def}
\end{equation}
where $\Delta(x)$ satisfies the normalization constraint
\begin{equation}
\int dx\, \Delta(x) = 0.
\label{eq:Delta_norm}
\end{equation}

The Kullback--Leibler divergence is defined as
\begin{equation}
D_{\mathrm{KL}}(P \| Q)
= \int dx\, P(x)\,\ln\frac{P(x)}{Q(x)}.
\label{eq:KL_def}
\end{equation}

Substituting Eq.~\eqref{eq:Q_def} into Eq.~\eqref{eq:KL_def}, we obtain
\begin{eqnarray}
D_{\mathrm{KL}}(P \| Q)
&=& \int dx\, P(x)\,\ln\left(\frac{P(x)}{P(x)+\Delta(x)}\right)\nonumber\\
&=& -\int dx\, P(x)\,\ln\!\left(1+\frac{\Delta(x)}{P(x)}\right).
\label{eq:log_expansion}
\end{eqnarray}

Assuming $\left|\Delta(x)/P(x)\right| \ll 1$, we expand the logarithm in a
Taylor series
$\ln(1+u) = u - \frac{u^2}{2} + \frac{u^3}{3} - \cdots$.
Retaining terms up to second order, Eq.~\eqref{eq:log_expansion}  gives
\begin{equation}
D_{\mathrm{KL}}(P \| Q)
\approx
-\int dx\, P(x)
\left[
\frac{\Delta(x)}{P(x)}
- \frac{1}{2}\left(\frac{\Delta(x)}{P(x)}\right)^2
\right].
\end{equation}

Distributing $P(x)$ inside the integrand,
\begin{equation}
D_{\mathrm{KL}}(P \| Q)
\approx
-\int dx\, \Delta(x)
+ \frac{1}{2}\int dx\, \frac{\Delta(x)^2}{P(x)}.
\label{eq:KL_split}
\end{equation}

By the normalization condition Eq.~\eqref{eq:Delta_norm},
the linear contribution in Eq.~\eqref{eq:KL_split} vanishes identically. Hence,
\begin{equation}
D_{\mathrm{KL}}(P \| Q)
\approx
\frac{1}{2}\int dx\, \frac{\Delta(x)^2}{P(x)}.
\label{eq:chi_square}
\end{equation}

For a fixed ordered pair of nodes $(i,j)$, the cumulative error accumulated
along the shortest path of length $d_{ij}$ arises from $d_{ij}$ independent
Bernoulli trials with bias $r$. Each trial contributes an error of magnitude
$\pm 1/(n-1)$. The mean and variance of the total error are therefore
\begin{eqnarray}
m_{ij}
&=&
(2r-1)\, d_{ij}\,\frac{1}{n-1}
\;\approx\;
(2r-1)\, c_{ij},
~~
0 < c_{ij} := \frac{d_{ij}}{n} < 1,
\nonumber\\
v_{ij}
&=&
4r(1-r)\, d_{ij}\,\left(\frac{1}{n-1}\right)^2
\;\approx\;
4r(1-r)\,\frac{c_{ij}}{n},
\nonumber
\end{eqnarray}
where the approximations hold in the large-network limit $n\to\infty$.
The variance $v_{ij}$ vanishes as $\mathcal{O}(1/n)$, implying that
fluctuations around the mean shift are negligible asymptotically.

If distances from node $j$ are approximately homogeneous across the network,
we may replace $c_{ij}$ by a node-dependent mean value $c_j$, yielding
$m_{ij}\approx m_j=(2r-1)c_j$.
Neglecting fluctuations of order $O(n^{-1/2})$, the perceived distribution at
node $j$ becomes
\begin{equation}
Q^{(j)}(x)
=
\frac{1}{n}\sum_i P(x-m_{ij})
\;\approx\;
P(x-m_j).
\end{equation}
Since $0<c_j<1$ and $|2r-1|\leq 1$, we have $|m_j|<1$.

In a mean-field description, the empirical distribution of rescaled distances
$c_j$ converges to a density $\rho(c)$ with finite moments.
Introducing $\alpha := 2r-1 \in [-1,1]$, we write $m_j=\alpha c_j$, with $|m_j|<1$.
Assuming self-averaging of distance statistics in the large-$n$ limit,
node-dependent moments converge to network-wide constants,
\[
\overline{m}_j \to \overline{m},
\qquad
\overline{m_j^2} \to \overline{m^2}.
\]
Accordingly, the perceived distribution becomes asymptotically independent
of $j$, and we write $Q_j(x)\approx Q(x)$.

The node-averaged perceived distribution is then
\begin{eqnarray}
Q(x)
&=& \frac{1}{n} \sum_j Q^{(j)}(x) =
\int_0^1 P\!\left(x-m(c)\right)\rho(c)\,dc
\nonumber\\
&=&
P(x)
- \overline{m}\,P'(x)
+ \frac{\overline{m^2}}{2}\,P''(x)
+ \mathcal{O}(m^3),
\end{eqnarray}
where $\overline{m^k}=\int_{c=0}^1 (\alpha c)^k\rho(c)\,dc$.
Therefore, $\Delta(x)=Q(x)-P(x)$ yields
\begin{equation}
\Delta(x)
=
-\overline{m}\,P'(x)
+
\frac{\overline{m^2}}{2}\,P''(x)
+
\mathcal{O}(m^3).
\end{equation}

For a Gaussian true distribution
$P(x)=\mathcal{N}(0,\sigma^2)$, the derivatives satisfy
\begin{equation}
\frac{P'(x)}{P(x)}=-\frac{x}{\sigma^2},
\qquad
\frac{P''(x)}{P(x)}=\frac{x^2}{\sigma^4}-\frac{1}{\sigma^2}.
\end{equation}

Using the quadratic approximation to the  KL divergence,
valid when $|\Delta(x)|\ll P(x)$, we obtain from \eqref{eq:chi_square}
\begin{eqnarray}
D_{\mathrm{KL}}(P\|Q)
&\approx&
\frac{1}{2}\int dx\,\frac{\Delta(x)^2}{P(x)}
\nonumber\\
&=&
\frac{\overline{m}^{\,2}}{2\sigma^2}
+
\frac{\overline{m^2}^{\,2}}{4\sigma^4}
+
\mathcal{O}(m^6).
\end{eqnarray}

Averaging over nodes yields the network-level misinformation
\begin{equation}
I
=
\frac{1}{n}\sum_j D_{\mathrm{KL}}(P\|Q^{(j)})
\;\approx\;
\frac{A}{2\sigma^2}
+
\frac{B}{4\sigma^4},
\end{equation}
where $A=\overline{m}^{\,2}$ and $B=\overline{m^2}^{\,2}$ depend on the
distance statistics of the network and satisfy $0<A,B<1$ for any connected
graph.

{\color{black}\section{Relation between R\'enyi and KL divergences in special limits}\label{app:B}

\begin{theorem}
Let $P(x)=\delta(x=0)$ be the Dirac probability measure peaked at $x=0$, and let $Q(x)$ be a probability measure such that
$Q(x=0)>0$,
so that $P\ll Q$. Then for all $\alpha>0$, $\alpha\neq 1$, the R\'enyi divergence satisfies $
D_{\alpha}(P\|Q) = D_{\mathrm{KL}}(P\|Q)$ 
and this equality holds for all $\alpha>0$ under the assumption $P\ll Q$. \textbf{Note:} The notation $P \ll Q$ between two probability distributions signifies that $P$ is absolutely continuous with respect to $Q$.

\end{theorem}

\emph{Proof:}
We follow the definition and conventions of R\'enyi divergence given in Erven et al.~\cite{erven_ieee_2014}. Since $P\ll Q$, the Radon--Nikodym derivative
$dP/dQ$ exists and is given by~\cite{shiryaev1996}
\begin{equation}\frac{dP(x)}{dQ(x)}
=
\begin{cases}
\dfrac{1}{Q(x=0)}, & x=0,\\[6pt]
0, & x\neq 0.
\end{cases}
\end{equation}

For all $\alpha>0$, $\alpha\neq 1$, the R\'enyi divergence is defined by integration
with respect to $Q$ as~\cite{erven_ieee_2014}
\begin{equation}
D_{\alpha}(P\|Q)
=
\frac{1}{\alpha-1}
\log
\int \left(\frac{dP}{dQ}\right)^{\alpha} dQ.
\end{equation}

Substituting the explicit form of $dP/dQ$, we obtain
\begin{equation}
\int \left(\frac{dP}{dQ}\right)^{\alpha} dQ
=
\int_{\{0\}}
\left(\frac{1}{Q(x=0)}\right)^{\alpha} dQ
=
Q(x=0)^{1-\alpha}.
\end{equation}

Therefore,
\begin{equation}
D_{\alpha}(P\|Q)
=
\frac{1}{\alpha-1}
\log\left(Q(x=0)^{1-\alpha}\right)
=
-\log Q(x=0).
\end{equation}

This expression is independent of $\alpha$.

Finally, the KL divergence is given by
\begin{eqnarray}
D_{\mathrm{KL}}(P\|Q)
&=&
\int \log\left(\frac{dP}{dQ}\right) dP 
=
-\log Q(x=0). \nonumber\\
\end{eqnarray}

Hence, $D_{\alpha}(P\|Q)=D_{\mathrm{KL}}(P\|Q)$ for all $\alpha>0$, $\alpha\neq 1$,
under the assumption $P\ll Q$.

\begin{theorem}
Let $P(x)$ and $Q(x)$ be smooth, strictly positive PDFs, then for any $\alpha>0$, $\alpha\neq1$, the R\'enyi
divergence satisfies
$ 
D_\alpha(P\|Q)
=
\alpha\, D_{\mathrm{KL}}(P\|Q)
+\mathcal{O}(\Delta^3)
$.  
\end{theorem}

\emph{Proof:}
The R\'enyi divergence of order $\alpha$ is defined as
\begin{equation}
 D_\alpha(P\|Q)
=
\frac{1}{\alpha-1}
\log
\int dx\, P(x)^\alpha Q(x)^{1-\alpha}.   
\end{equation}

Substituting $Q(x)=P(x)+\Delta(x)$ gives
\begin{equation}
 D_\alpha(P\|Q)
=
\frac{1}{\alpha-1}
\log
\int dx\, P(x)
\left(1+\frac{\Delta(x)}{P(x)}\right)^{1-\alpha}.   
\end{equation}

Using the expansion

$ (1+u)^{1-\alpha}
=
1+(1-\alpha)u
-\frac{\alpha(1-\alpha)}{2}u^2
+\mathcal{O}(u^3)$, 
we obtain

\begin{eqnarray}
&&\int dx P(x)
\left(1+\frac{\Delta(x)}{P(x)}\right)^{1-\alpha} \nonumber\\
&=&\int dx
\Big(
P(x)
+(1-\alpha)\Delta(x)
-\frac{\alpha(1-\alpha)}{2}\frac{\Delta(x)^2}{P(x)}
\nonumber\\&&+\mathcal{O} \left(  \frac{\vert\Delta(x)\vert^3}{P(x)^2}  \right)
\Big).
\end{eqnarray}

The linear term vanishes due to normalization,
$\int dx~\Delta(x)=0$. Therefore the above expression becomes
\begin{eqnarray}
&&\int dx~ P(x)
\left(1+\frac{\Delta(x)}{P(x)}\right)^{1-\alpha} \nonumber\\
&=&
1
-\frac{\alpha(1-\alpha)}{2}
\int dx \left[\frac{\Delta(x)^2}{P(x)}
+\mathcal{O}\left(  \frac{\vert\Delta(x)\vert^3}{P(x)^2}  \right) \right].
\label{eq: B9}
\end{eqnarray}

Taking the logarithm of \ref{eq: B9} and expanding the logarithm in a
Taylor series, and retaining terms up to $|\Delta|^2$, we obtain
\begin{equation}
D_\alpha(P\|Q)
=
\frac{\alpha}{2}
\int dx\,\frac{\Delta(x)^2}{P(x)}
+\mathcal{O}\left( \int dx~  \frac{\vert\Delta(x)\vert^3}{P(x)^2}  \right).
\label{B10}
\end{equation}

On the other hand, from \eqref{eq:chi_square}
\begin{equation}
D_{\mathrm{KL}}(P\|Q)
=
\frac{1}{2}
\int dx\,\frac{\Delta(x)^2}{P(x)}
+\mathcal{O}\left( \int dx~  \frac{\vert\Delta(x)\vert^3}{P(x)^2}  \right).
\label{B11}
\end{equation}

Comparing the two expressions \eqref{B10} and \eqref{B11} we show 

\begin{equation}
 D_\alpha(P\|Q)
=
\alpha D_{\mathrm{KL}}(P\|Q)
+\mathcal{O}\left( \int dx~  \frac{\vert\Delta(x)\vert^3}{P(x)^2}  \right).   
\end{equation}}

\end{document}